\documentclass[reprint,amssymb, amsmath, aip,cha,twocolumn]{revtex4-2}
\usepackage{graphicx}
\usepackage{color}
\usepackage{epsfig}
\usepackage{ifpdf}
\usepackage{url}%
\usepackage{amsmath}
\usepackage{array}
\usepackage{float}
\usepackage{bm}
\usepackage{subcaption}
\usepackage[colorlinks=true,linkcolor=blue]{hyperref}%
\usepackage{cleveref}
\usepackage{ulem}
\expandafter\ifx\csname package@font\endcsname\relax\else
\expandafter\expandafter
\expandafter\usepackage
\expandafter\expandafter
\expandafter{\csname package@font\endcsname}%
\fi
\begin{document}
	\title{Spatio-temporal dynamics of anisotropic emission from nano second laser produced aluminium plasma}%
	\author{Geethika B R}%
	\email{geethika.br@ipr.res.in}
	\affiliation{Institute For Plasma Research, Bhat, Gandhinagar, Gujarat, 382428, India}%
	\affiliation{Homi Bhabha National Institute, Training School Complex, Anushaktinagar, Mumbai, 400094, India}%
	\author{Jinto Thomas}
	\email{jinto@ipr.res.in}
	\affiliation{Institute For Plasma Research, Bhat, Gandhinagar, Gujarat, 382428, India}%
	\affiliation{Homi Bhabha National Institute, Training School Complex, Anushaktinagar, Mumbai, 400094, India}%
	\author{Milaan Patel}
	\affiliation{Institute For Plasma Research, Bhat, Gandhinagar, Gujarat, 382428, India}%
	\affiliation{Homi Bhabha National Institute, Training School Complex, Anushaktinagar, Mumbai, 400094, India}%
	\author{Renjith Kumar R}
	\affiliation{Institute For Plasma Research, Bhat, Gandhinagar, Gujarat, 382428, India}%
	\affiliation{Homi Bhabha National Institute, Training School Complex, Anushaktinagar, Mumbai, 400094, India}%
	
	\author{Hem Chandra Joshi}
	\affiliation{Institute For Plasma Research, Bhat, Gandhinagar, Gujarat, 382428, India}%

	\date{\today}
	\begin{abstract}
		Polarized emission carries captivating information and can help understand various elementary processes involving collisions within the plasma as well as in radiative transitions. In this work, we investigate the spatio-temporal dependence of the emission anisotropy of a nanosecond laser produced aluminium plasma at 100 mbar background pressure. We observe that the anisotropy of the emission spectra exhibits interesting spatio-temporal characteristics which in turn depend on the charge state of the emitting species. The degree of polarization (DOP) is found to reverse its sign along the plume propagation direction.  Observed behaviour in DOP appears to be due to the contribution from various involved atomic processes. However, closer to the sample the contribution from the self-generated magnetic field predominantly affect the polarization. On the other hand, the effect of the self generated magnetic field on the observed polarized emission is insignificant as the plume propagates away from the sample. This is of particular interest in polarization resolved laser induced breakdown spectroscopy as spatio-temporal profile of the degree of polarization has to be properly taken into account prior to the spectral analysis.
	\end{abstract}
	\maketitle
	\section{Introduction}\label{sec:intro}
	
	Anisotropy, in general is referred to the property of exhibiting different behaviour of a particular parameters when measured along different directions. In laser produced plasma, it has been observed that certain emission lines have preferential  directional enhancement resulting in polarized emission. Polarized emission from laser produced plasma (LPP) has been the subject of investigation for many groups in the past\cite{kieffer1992,wubetu,SHARMA20073113,aghababaei,ASGILL20101033,spatial_pps,Labutin2016}. Anisotropy in emission spectra of the expanding plasma plume is, naturally, riveting due to its importance in deciphering electron distribution\cite{kieffer1992} and also extracting information regarding self-generated electric\cite{PhysRevE_Jinto} and magnetic fields\cite{SHARMA20073113}. Moreover, it is an interesting aspect in the development of extreme ultraviolet (EUV) light source for nano-lithography\cite{lucas_euv}. Polarization of light emission from the plasma has found applications in enhancing the signal to noise
	ratio in Laser Induced Breakdown Spectroscopy (LIBS), termed as polarization resolved LIBS (PRLIBS)\cite{aghababaei,TakashiFujimoto_1999,Zhao_2014} \par
	Kieffer et al\cite{kieffer1992} observed anisotropic emission from He-like emission from laser produced plasma for the first time and ascribed it to the anisotropy in the electron velocity distribution.
	Fujimoto et.al\cite{TakashiFujimoto_1998} explained the anisotropy in emission from the fact that, when an atom is excited by an electron, it maintains the electron's direction and behaves as a dipole that oscillates in the collision direction, producing dipole radiation.
	Sharma et.al \cite{SHARMA20073113} observed polarized emission of aluminium at atmospheric pressure and attributed this to the possible presence of self-generated magnetic field due to Rayleigh Taylor instability.Similar study \cite{NAGLI} proposed that self-generated magnetic field induced by the pump beam is responsible for polarization effects. Wubetu et al \cite{Wubetu_2020} studied the polarized emission from aluminium and copper plasma at low background pressure and showed that the anisotropic nature of emission has some dependence on the background pressure. They have suggested that anisotropy in the plasma plume should be driven by radiative recombination during the early phase of the plume. 
	On the other hand, in another study by Asgill et.al \cite{ASGILL20101033}, they did not notice anisotropy in emissions from gaseous samples and only a small polarization was observed for the solid samples at atmospheric pressures. They observed that the polarization was spectrally flat in all cases, showing no significant contribution from the continuum emission compared to atomic emission. Aghababaei et al.\cite{AghababaeiNejad2017} also reported similar observation regarding atomic and continuum emissions at atmospheric pressure. 
	\par
	Kim et al\cite{spatial_pps} observed variation in the extent of polarization with distance from the target material. They ascribed the possible reason as recombination from the higher ionic states. In some earlier studies\cite{PhysRevE_Jinto,asymmetry}, strong spectral asymmetry was reported for the neutral emission. The observed asymmetry was attributed to the presence of a large micro electric field in the laser produced plasma depending on time and spatial location within the plasma plume. Apart from these, few studies \cite{Hammond1989,Wolcke_1983} suggested reversal in the preferred polarization state of the emission based on the energy of the electrons involved in the excitation process. They further showed that energy of electrons can affect the sign of degree of polarization of the emission.
	\par
	From these reports, it can be inferred that the spatio-temporal dynamics of the polarized emission is less explored. The present work deals with a detailed study on the spatio-temporal evolution of anisotropic emission from various charge states and spectral lines in laser produced aluminium plasma using a nano-second(ns) laser with comparatively lower laser energy and at a moderately high background pressure. It reveals an interesting feature of flip in the polarized emission along the plume expansion direction. Possible scenarios concerning anisotropic emission are discussed. 
	
	\section{Experimental Set-up}\label{sec:setup}
	
	\begin{figure}[h!]
		\includegraphics[scale = 0.3,trim = {0.8cm 1.5cm 0 0}, clip]{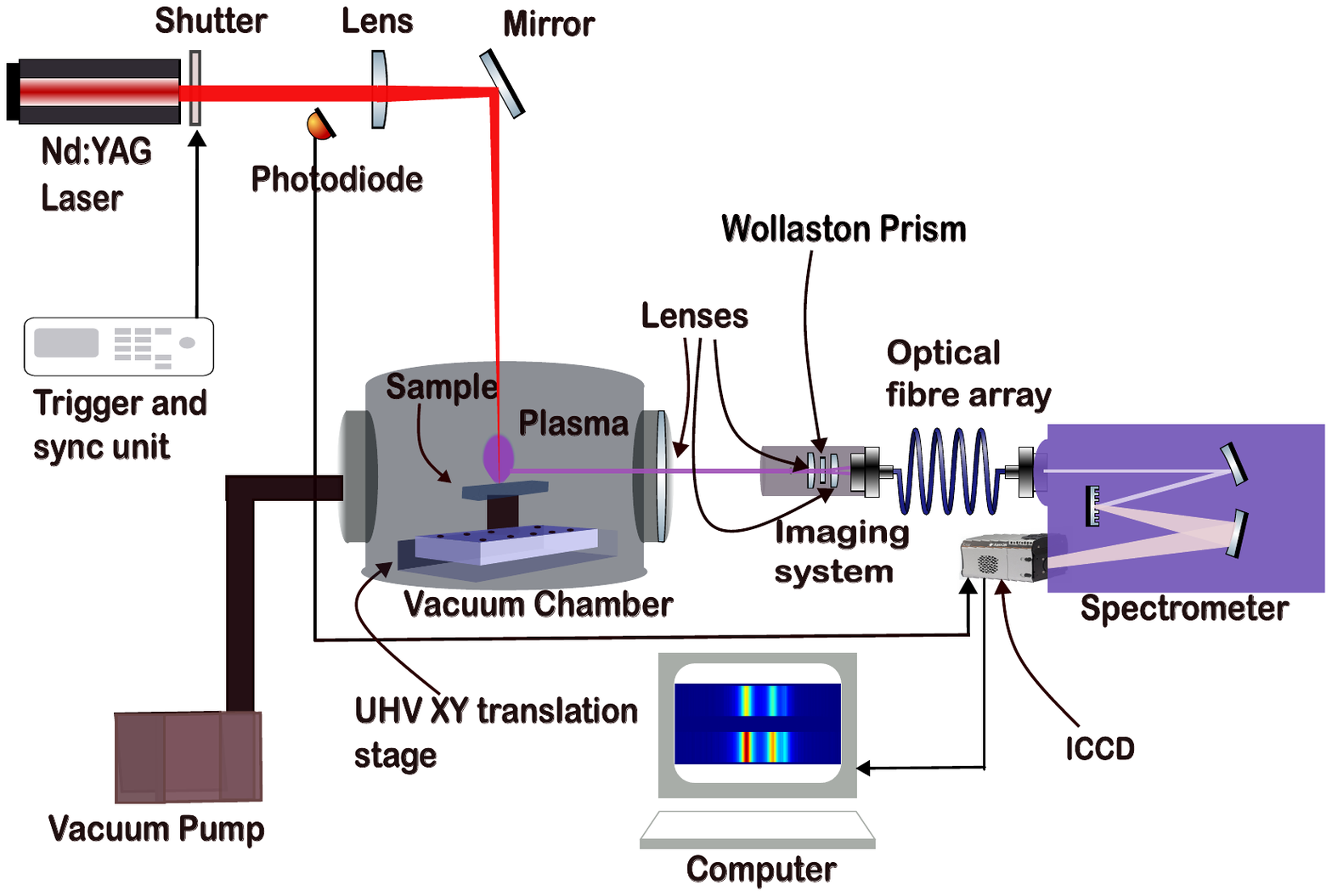}
		\caption{\label{fig:expsetup} A schematic diagram of experimental setup showing the arrangement of the sample inside the vacuum chamber, Nd:YAG laser used for the ablation, imaging system, spectrometer and ICCD. The vacuum pump shown is a dry pump which gives base pressure of approximately $10^{-2}$mbar. Trigger and sync unit ensures the opening of the shutter to regulate the number of pulses from the laser. }
	\end{figure}
	A schematic diagram of the experimental setup is shown in figure \ref{fig:expsetup}. 
	It consists of an upright cylindrical vacuum chamber pumped by a dual stage scroll pump (160 L/m). The chamber is first evacuated to $10^{-2}$ mbar and then filled with nitrogen gas up to 100 mbar which is maintained during the experiment.
	Aluminium (Al) is used as the target material with dimensions 50mm$\times$50mm$\times$5mm kept on an ultra-high vacuum compatible XY translational stage which ensures a new sample position for each ablation. Nd:YAG laser at its fundamental wavelength 1064 nm with pulse width 10 ns, energy 150 mJ and repetition rate 30 Hz was used for the experiments. The laser beam was focused on the sample using a 600 mm plano-convex lens positioned such that to form a  spot diameter of $\approx$ 1 mm on the target. \par
	
	The optical emission from the plasma was collected using an imaging system with unit magnification and a spatial resolution of 1 mm. The first lens of the imaging system is kept inside the vacuum chamber at 2$f$ distance from the plasma plume to increase the light collection. A Wollaston prism (angle of deviation $1^o$) is used to separate horizontal (H) and vertical (V) polarizations of emissions. 
	A polarized helium neon laser is used to ascertain the H and V polarizations.
	An optical fiber array (each fibre having a diameter of 600 microns) kept at the focal plane of the imaging system collects the H and V polarization on different fibers.The imaging system forms a well separated ($\sim$ 2 mm in this case of magnification 1) images of H and V polarizations at the image plane. The fiber array is coupled to McPhereson model 2061, 1 meter Czerny Turner spectrometer with 1200g/mm grating (0.06 nm resolution) through an F\/ \# matching optical system to record the polarization-resolved emission spectra. An intensified charge coupled device (Andor - ICCD) coupled with the spectrometer is gated for 30 ns and synchronized with a trigger using a fast photo diode.
	This arrangement avoids uncertainties arising from shot to shot variations, laser intensity fluctuations, uncertainty in quantum efficiency\cite{exp_set} and gains of two different detectors.
	Gate delay of the ICCD is varied to record the temporal evolution of the spectral emission. The imaging system, optical fiber array and spectrometer are calibrated with few transition lines of neon spectral calibration lamp to confirm spectral accuracy as well as the polarization independence of the imaging and measurement systems. The observed ratio of H and V polarization intensities is 0.99 $\pm$ 0.05, showing that the entire setup is polarization independent. Intensity calibration of the spectrograph along with the imaging system is performed using a white-light source kept at the object plane of the imaging system.
	All of the emission data are averaged for over 20 ablations to reduce the statistical variations in recorded intensity. The statistical variation in the recorded intensity is well within 5\%. The imaging lens system is mounted on a precision translational stage to scan the emission along the propagation axis of the plasma plume.

	\section{Results and Discussion}\label{sec:results}

	\begin{figure*}[!tbp]
		\includegraphics[scale=0.42,trim = {0.5mm 0 2.5cm 0},clip]{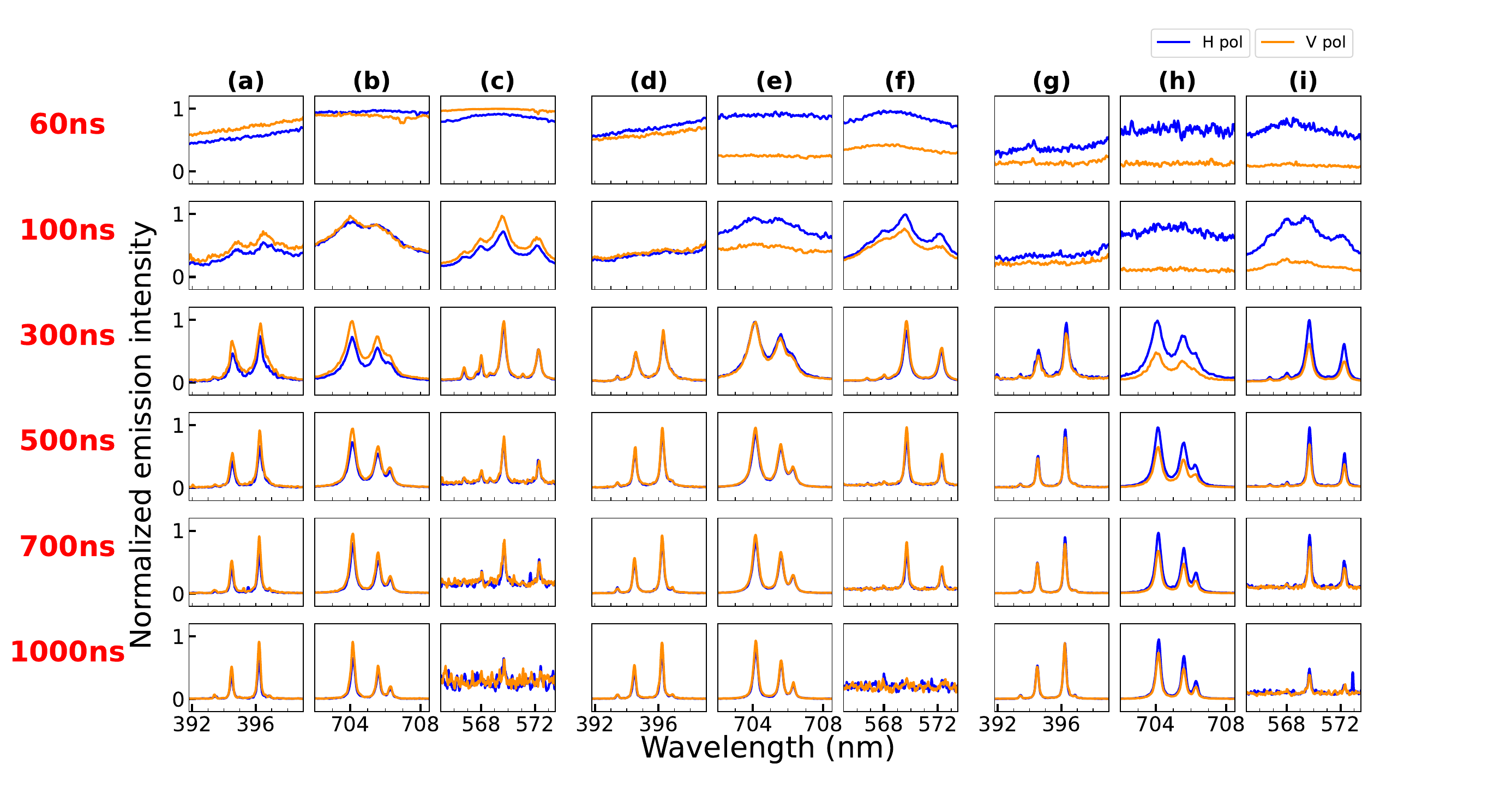}
		\caption{\label{fig:diffwave_all_mm}Temporal evolution of the intensities of horizontal (blue) and vertical (orange) polarizations of different wavelengths. The first three columns, \textbf{a,b,c} shows the emission from Al I- 396.15 nm, Al II- 704.21 nm and Al III-569.66 nm respectively at 1.5mm away from the sample, the next three columns (\textbf{d,e,f}) have the spectra of the same ionic states at 2.5mm away from the sample and the last three columns (\textbf{g,h,i}) are at 3.5 mm away from the sample. All plots are normalised to the maximum intensity of emission among the two polarizations for better visibility. }
	\end{figure*}
	\begin{figure*}[]
		\includegraphics[scale=0.41,trim = {0 0 2cm 0},clip]{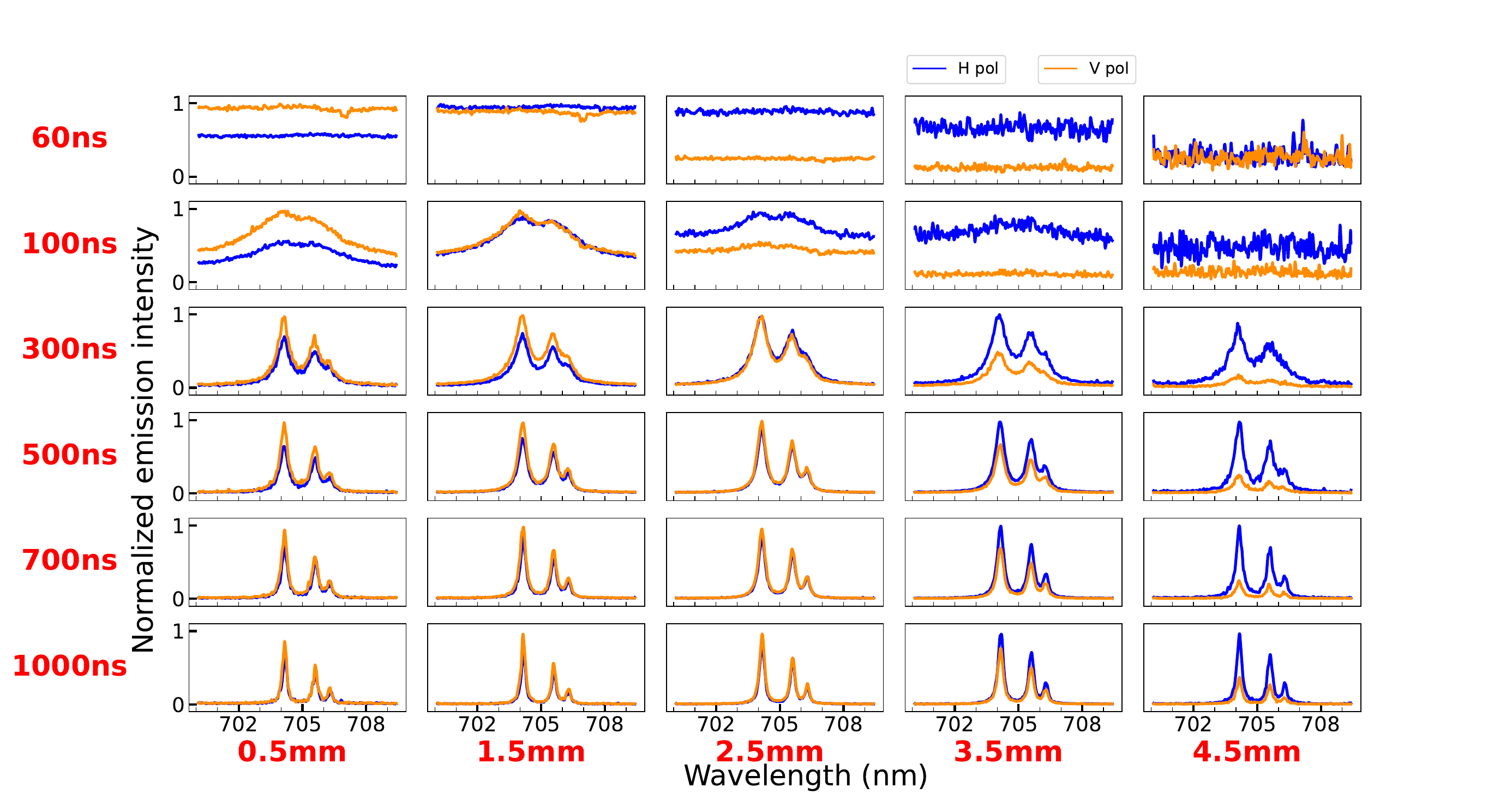}
		\caption{\label{fig:704_mod} Spatio-temporal evolution of the emission spectrum of Al II. Row-wise it shows the spectra with increasing time after the laser ablation and column-wise the spectra with increase in the distance from the sample/target. Blue curve shows the horizontal polarization and the orange shows the vertical polarization. All plots are normalised to the maximum intensity of emission among the two polarizations.  }
	\end{figure*}
	Figure~ \ref{fig:diffwave_all_mm} shows the temporal evolution (60 ns to 1000 ns) of polarization-resolved spectra of three charge states of Al (Al I- 396.15 nm, Al II- 704.21 nm and Al III-569.66 nm) at three different spatial locations (1.5 mm, 2.5 mm, and 3.5 mm) in the plasma plume. Intensity of each spectrum is normalized with the maximum value among the polarizations for easy comparison.
	At a distance of 1.5 mm from the sample (columns \textbf{(a),(b),(c)} of figure \ref{fig:diffwave_all_mm}), polarization resolved spectra show an interesting trend in time evolution depending on the charge states. For instance, at the beginning of the plume expansion, the spectra corresponding to the emission from all charge states dominates from continuum and have minimal intensity difference between the two polarizations. 
	A tiny dip in the spectra of continuum emission appears to be due to the artefact in camera pixels. Line emission emerges as time increases to 100-300 ns range with slight polarization. It is to be noted that Al III shows trace of anisotropy at 100 ns with dominant V polarization, which subsequently disappears at 300 ns, whereas in the case of Al II, anisotropy is not observed till 100 ns and appears at 300 ns with a dominant V polarization. In case of Al I, V polarization is dominant till 300 ns. At later times, anisotropy appears minimal irrespective of the charge states.
	\par
	
	When the distance is increased to 2.5 mm, (columns \textbf{(d),(e),(f)} of figure \ref{fig:diffwave_all_mm}), observed anisotropy for continuum is increased substantially for wavelength range centred at 570 nm and 706 nm in the initial time of the expansion, but with flipped polarization. This flip within a small distance essentially demands a precise imaging system, as used here, which prevents the averaging of intensities from adjacent locations. 
	Until 100 ns, the wavelength ranges previously mentioned are dominated by H polarization, thereafter, no substantial anisotropy is observed for any charge state. Further, at 3.5 mm from the sample (columns \textbf{(g),(h),(i)} of figure \ref{fig:diffwave_all_mm}) intensity for H polarization remains higher but anisotropy is observed for relatively longer duration prominently for Al II emission.
	To conclude, figure \ref{fig:diffwave_all_mm} shows that the emission from the plasma exhibits anisotropy depending on the charge state, space, as well as time. 
	\par
	
	\begin{table}[ht]
		\caption{Table shows the detailed spectral parameters of the observed emission lines from different ionic states of Al. g$_k$ represents the statistical weight and A$_{ki}$ is the transition probability. E$_k$ represents the energy of upper state involved in the transition. (from NIST database\cite{NIST_ASD})}
		\centering
		\begin{tabular}{ | p{1cm} | p{1.6cm} | p{1.5cm} | p{2.5cm} |p{1.5cm} |}
			\hline
			Ionic state & Wavelength (nm)  & g$_k$A$_{ki}$$\times$10$^8$ (s$^{-1}$) & Spectral Terms of Transition & E$_k$ (eV)\\ \hline
			Al I & 394.40 & 0.998  & $^2$S$_{1/2}$ $\rightarrow$ $^2$P$_{1/2}$ & 3.1427 \\ \hline
			Al I & 396.15 & 1.97  & $^2$S$_{1/2}$ $\rightarrow$ $^2$P$_{3/2}$ & 3.1427 \\ \hline
			Al II & 559.33 & 4.63  & $^1$D$_{2}$ $\rightarrow$ $^1$P$_{1}$ & 15.4725 \\ \hline
			Al II & 623.17 & 4.20  & $^3$D$_{2}$ $\rightarrow$ $^3$P$_{1}$ & 15.0620 \\ \hline
			Al II & 624.34 & 7.77  & $^3$D$_{3}$ $\rightarrow$ $^3$P$_{2}$  & 15.0620 \\ \hline
			Al II & 704.21 & 2.89  & $^3$P$_{2}$ $\rightarrow$ $^3$S$_{1}$ & 13.0767 \\ \hline
			Al II & 705.66 & 1.72  & $^3$P$_{1}$ $\rightarrow$ $^3$S$_{1}$ & 13.0730 \\ \hline
			Al II & 706.36 & 0.573  & $^3$P$_{0}$ $\rightarrow$ $^3$S$_{1}$ & 13.0713 \\ \hline
			Al III & 452.92 & 14.9  & $^2$D$_{5/2}$ $\rightarrow$ $^2$P$_{3/2}$ & 20.5548 \\ \hline
			Al III & 569.66 & 3.51  & $^2$P$_{3/2}$ $\rightarrow$ $^2$S$_{1/2}$ & 17.8182 \\ \hline
			Al III & 572.27 & 1.73  & $^2$P$_{1/2}$ $\rightarrow$ $^2$S$_{1/2}$ & 17.8083 \\ \hline
		\end{tabular}
		\label{table:em_param}
	\end{table}
	
	Previously reported \cite{wubetu,SHARMA20073113,Wubetu_2020} studies on anisotropic emissions were mostly restricted to the temporal evolution for a few transitions. Kim et.al \cite{spatial_pps} studied spatial variation of anisotropy, however, limited to a much smaller distance and for rather low laser fluence. They used Al III emission at 572.27 nm to calibrate the setup assuming there was no anisotropy for this particular transition. However, in our observation, this transition also exhibits considerable anisotropy (figure \ref{fig:diffwave_all_mm}). Here we would like to point out that the present results show explicit spatial dependence on anisotropy. This observation has ramifications, especially when the polarization is exploited to enhance the signal-to-noise ratio in LIBS. In such a scenario, the polarization of signal has to be optimized after studying its spatial and temporal information to get the desired results. 

	\par
	
	For better understanding of spatio-temporal evolution of anisotropy, Al II line emissions peaking at wavelength 704.21 nm, 705.66 nm and 706.36 nm were recorded up to 5 mm from the sample with 1 mm spatial resolution.
	Figure~\ref{fig:704_mod} shows the polarization-resolved emission spectra of Al II lines at different delay times and positions. It can be noticed that the anisotropy depends significantly on the spatial location and the delay time. At 60 ns, strong evidence of anisotropic emission is observed for continuum, which eventually shows the emergence of line emission from the background as the time increases to 300 ns. It is important to note that the polarization reverses (flips) as we move away from the sample. For instance, in case of 60ns and 100 ns, near the sample (at a distance of 0.5 mm), emission with V polarization dominates substantially. At 1.5 mm, emission intensity for these two delay times almost remains the same, but as the distance increases, the emission intensity for H polarization dominates. As the time increases, the switching point of polarization changes from 1.5 mm to 2.5 mm and remains the same for the rest of the duration of the recorded spectra. It is interesting to note that the anisotropy of the Al II emission increases substantially at longer distances.	\par
	
\begin{figure}[h!]
		\includegraphics[scale=0.2,trim = {2cm 0.3cm 3cm 1cm},clip]{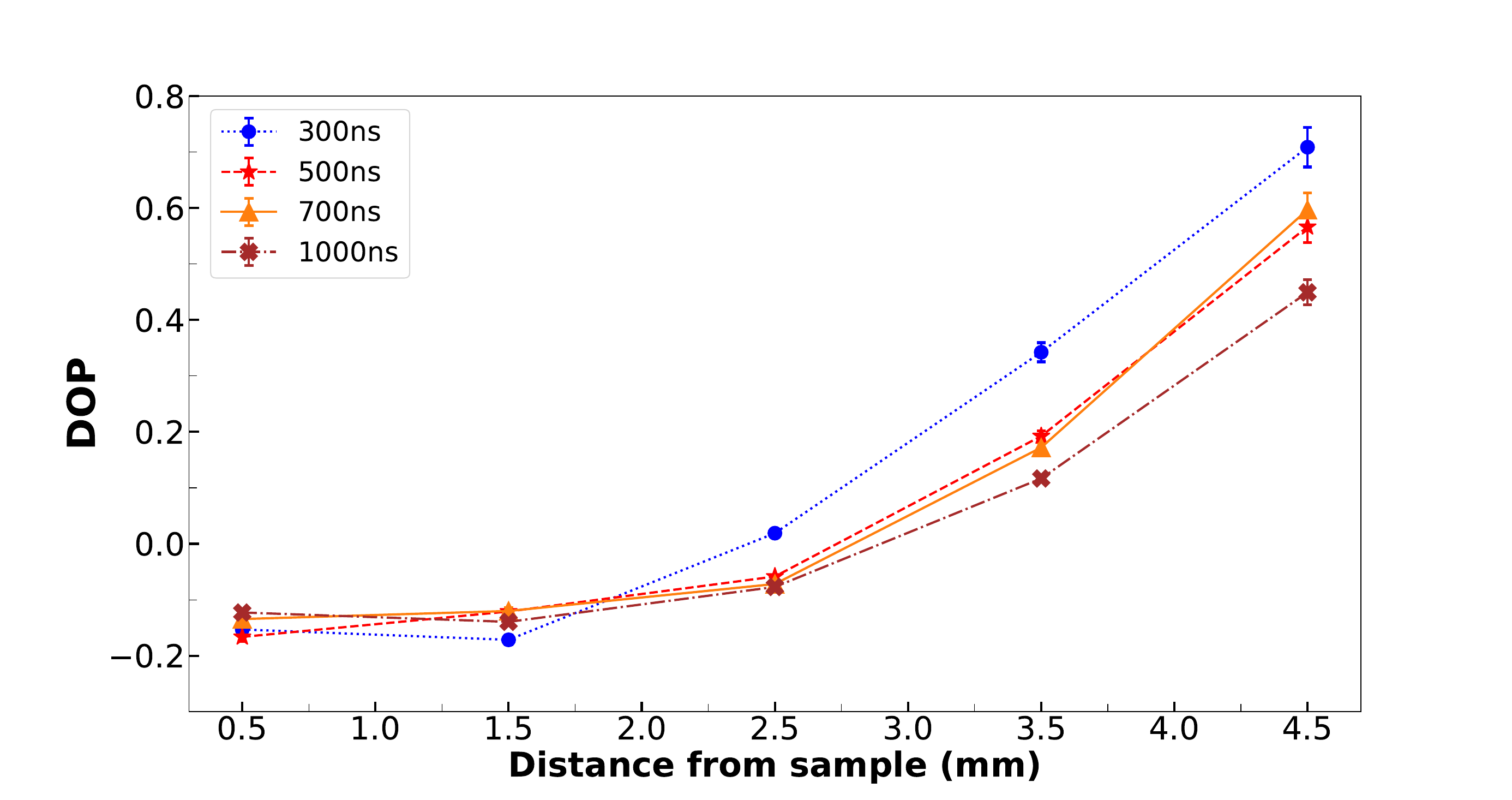}
		\caption{\label{fig:dop_704_mod}Spatial evolution of the DOP of Al II line of peak wavelength 704.2nm at different times after laser ablation. Each point is an average of 20 accumulations.}
	\end{figure}

	Anisotropy in the emission spectra can be quantified in terms of degree of polarization (DOP). DOP for a particular spectral emission can be estimated by the following relation\cite{wubetu}
	\begin{eqnarray}
		DOP = \frac{I_H-I_V}{I_H+I_V}
		\label{eq:dop}
	\end{eqnarray}
	where $I_H$ and $I_V$ are the intensities of H and V polarizations respectively. Figure~\ref{fig:dop_704_mod} shows the variation of DOP for Al II line emissions at various distances from the sample along the plume propagation axis for different delay times. This figure shows that the estimated DOP in the present work for the line emission is higher in comparison to the previously reported\cite{wubetu,spatial_pps}. Moreover, it reflects a clear reversal in sign at longer distances from the sample. To the best of our knowledge, this is the first time such a DOP reversal and its spatial dependence is reported.
	\par
	
	\begin{figure}[h!]
		\includegraphics[scale=0.2,trim = {2cm 0.3cm 3cm 1cm},clip]{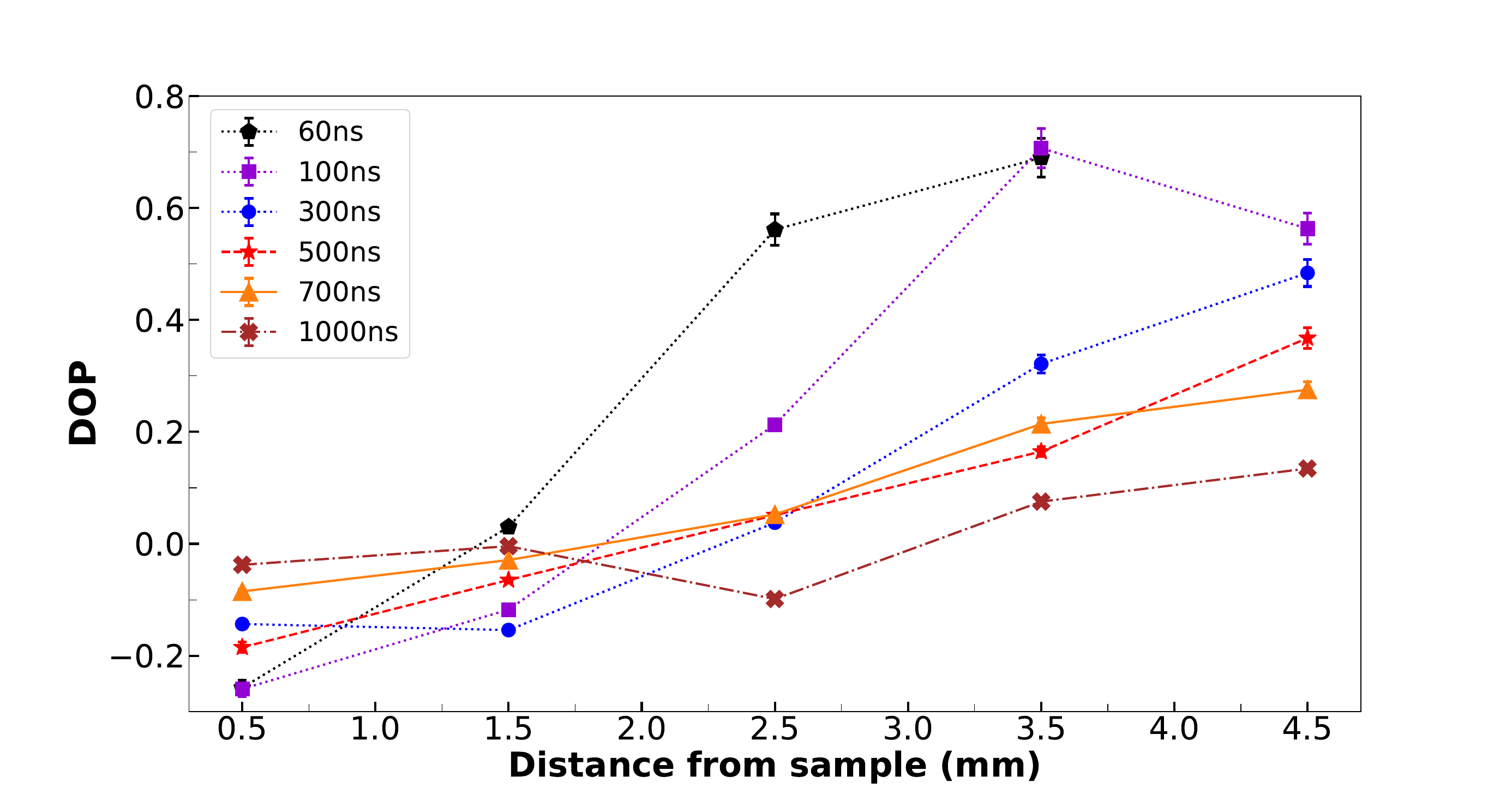}
		\caption{\label{fig:dop_cont_704_nog_mod}Spatial evolution of the DOP of continuum emission at different times after laser ablation. The value of continuum emission in range 700 to 709 nm was considered. }
	\end{figure}
\par
Similarly, the DOP of the continuum was also estimated for the range 700 to 709 nm. Since the recorded spectra are the convolution of different emission lines and the continuum, a multi peak fit was performed on the emission spectra to estimate the contribution of the continuum in the given wavelength range. A custom data processing algorithm is used for least squares fit of the required number of Lorentzian peaks to a given spectrum. The dark counts of the spectra recorded prior to the experiment are subtracted before processing the algorithm, so that the residual from the sum of all Lorentzian peaks can be taken as the continuum emission. The intensities of the continuum for the H and V polarizations were determined separately and DOP are estimated from equation~\eqref{eq:dop}. Figure~\ref{fig:dop_cont_704_nog_mod} shows the estimated DOP for the continuum at different spatial locations and times. It can be seen that similar to the line emission, the DOP for the continuum also shows flipping in the polarization. However, at longer distances there is a decrease in DOP of the continuum with time. \par

\begin{figure}
	\begin{minipage}{0.25\textwidth}
		\centering
		\includegraphics[width=0.8\linewidth,trim = {0cm 10cm 0cm 2cm},clip]{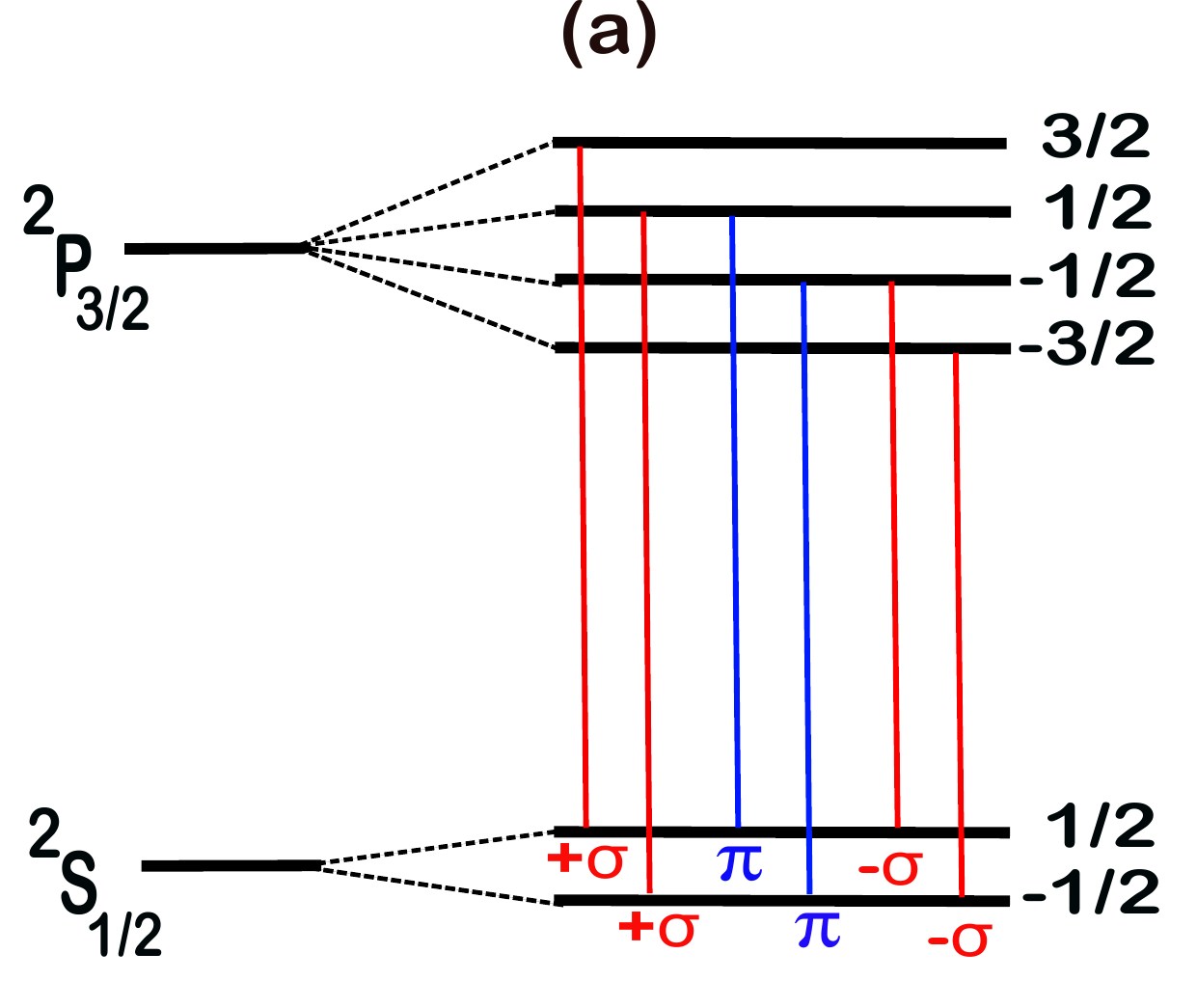}
		
		\label{fig:sub1}
	\end{minipage}%
	\begin{minipage}{0.25\textwidth}
		\centering
		\includegraphics[width=0.8\linewidth,trim = {0cm 10cm 0cm 2cm},clip]{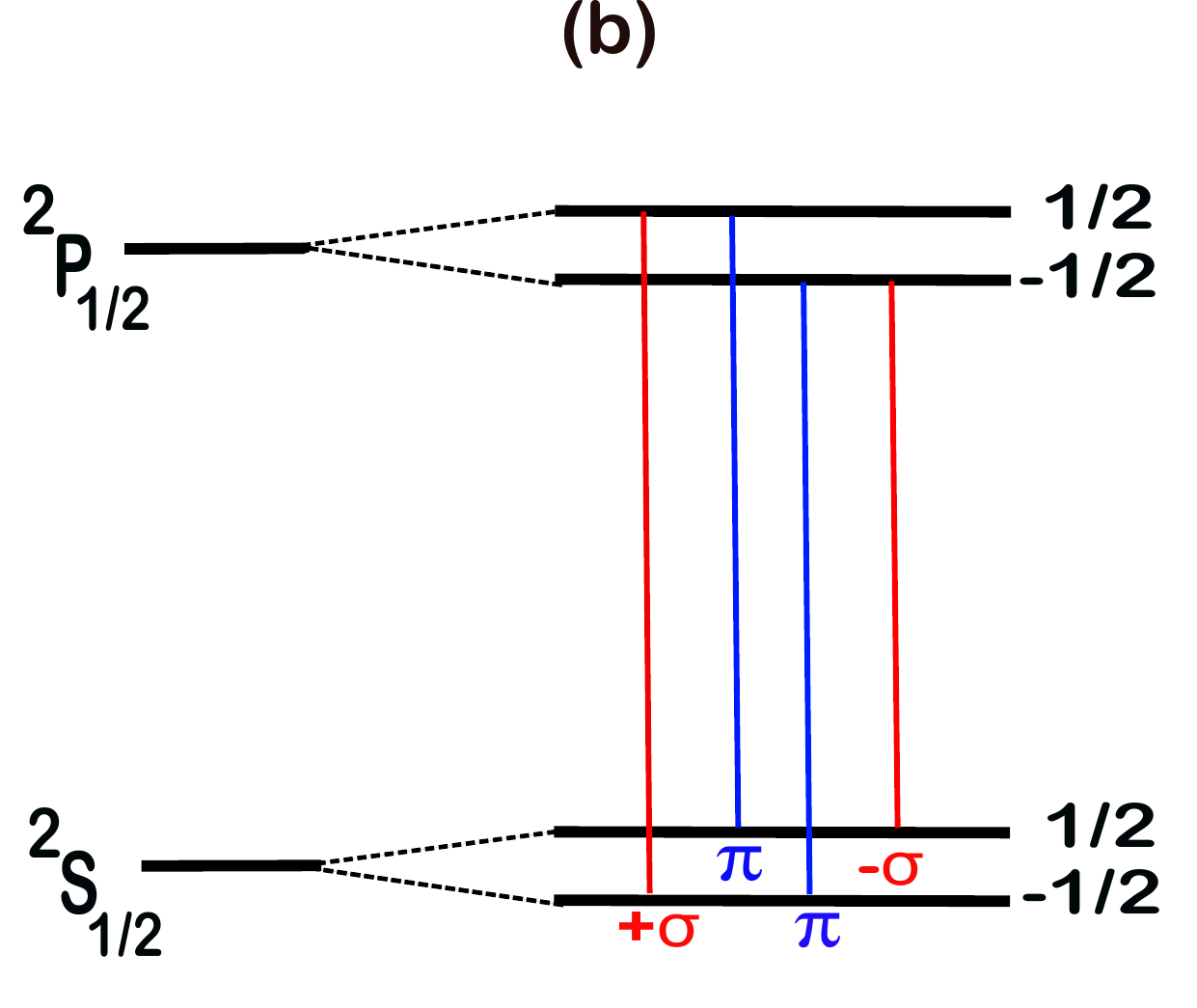}
		
		\label{fig:sub2}
	\end{minipage}
	\hfill
	\begin{minipage}{0.25\textwidth}
		\centering
		\includegraphics[width=0.8\linewidth,trim = {0cm 9cm 0cm 2cm},clip]{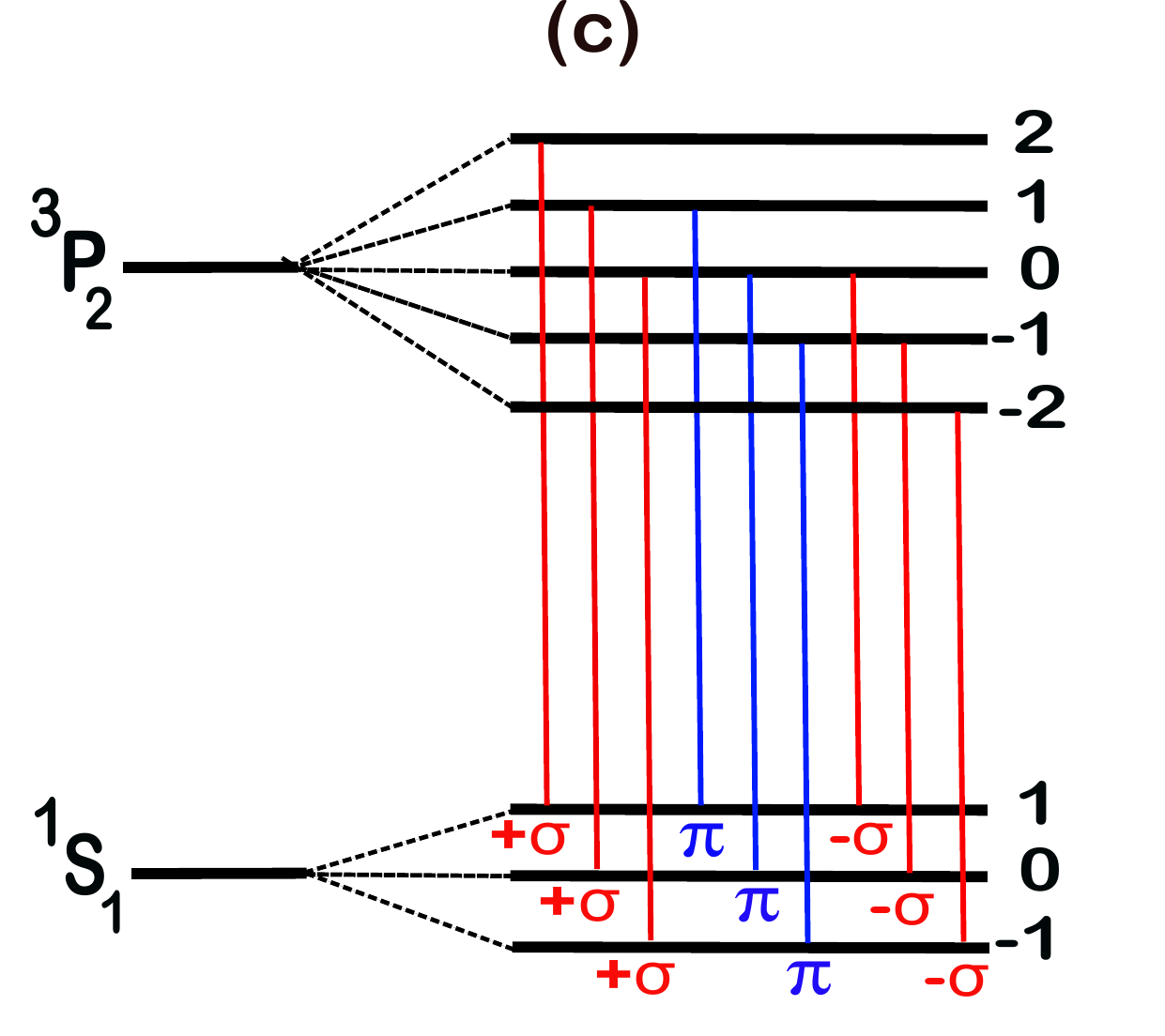}
		
		\label{fig:sub3}
	\end{minipage}%
	\begin{minipage}{0.25\textwidth}
		\centering
		\includegraphics[width=0.8\linewidth,trim = {0cm 9cm 0cm 2cm},clip]{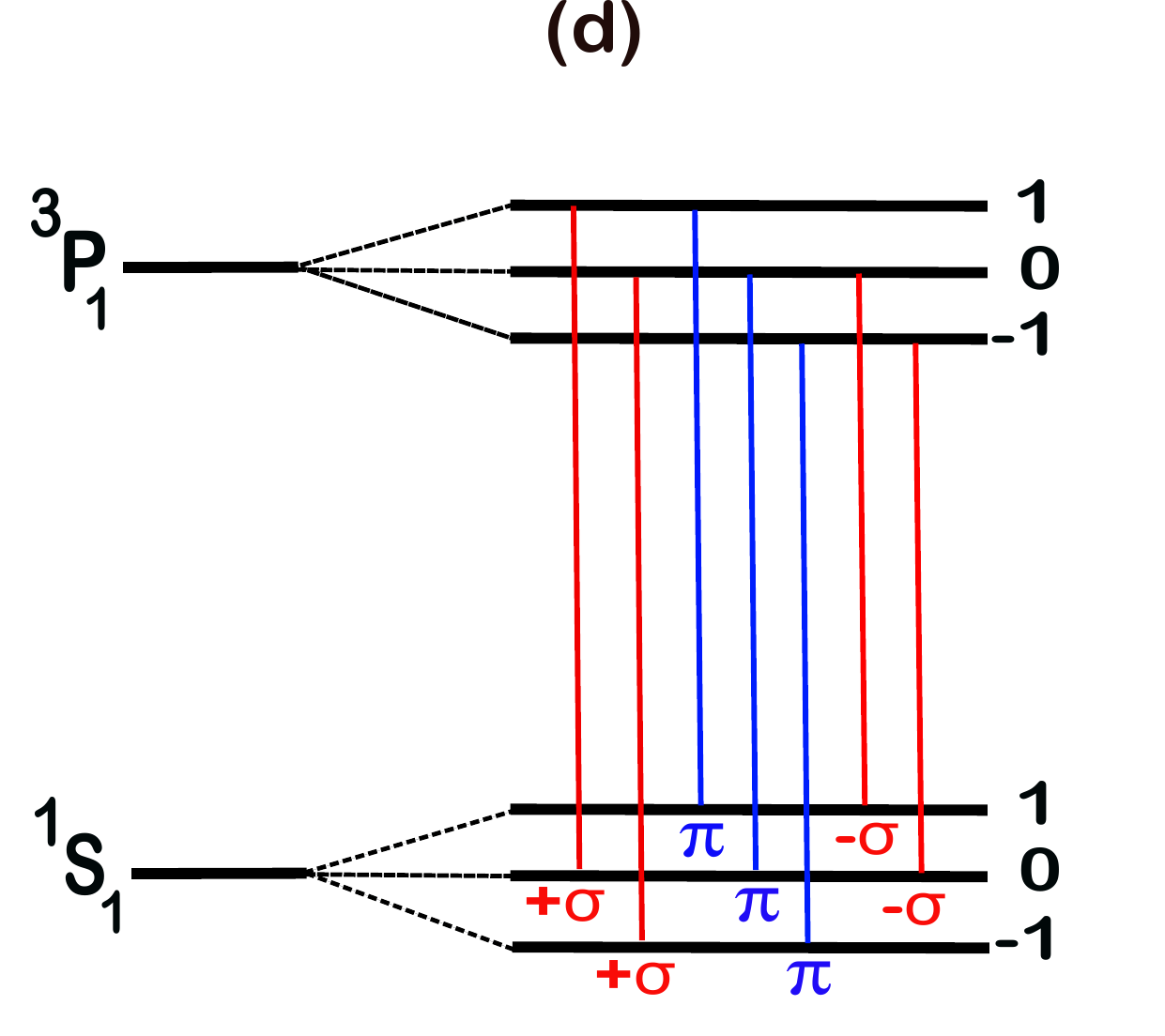}
		
		\label{fig:sub4}
	\end{minipage}
	\caption{Line components included in line transitions of (a) 569.66 nm (b)572.27 nm (c) 704.21 nm (d) 705.66 nm.}
	\label{fig:Jvalues}
\end{figure}

\begin{figure}[hbt]
	\includegraphics[scale=0.22,trim = {2cm 0 4.5cm 0},clip]{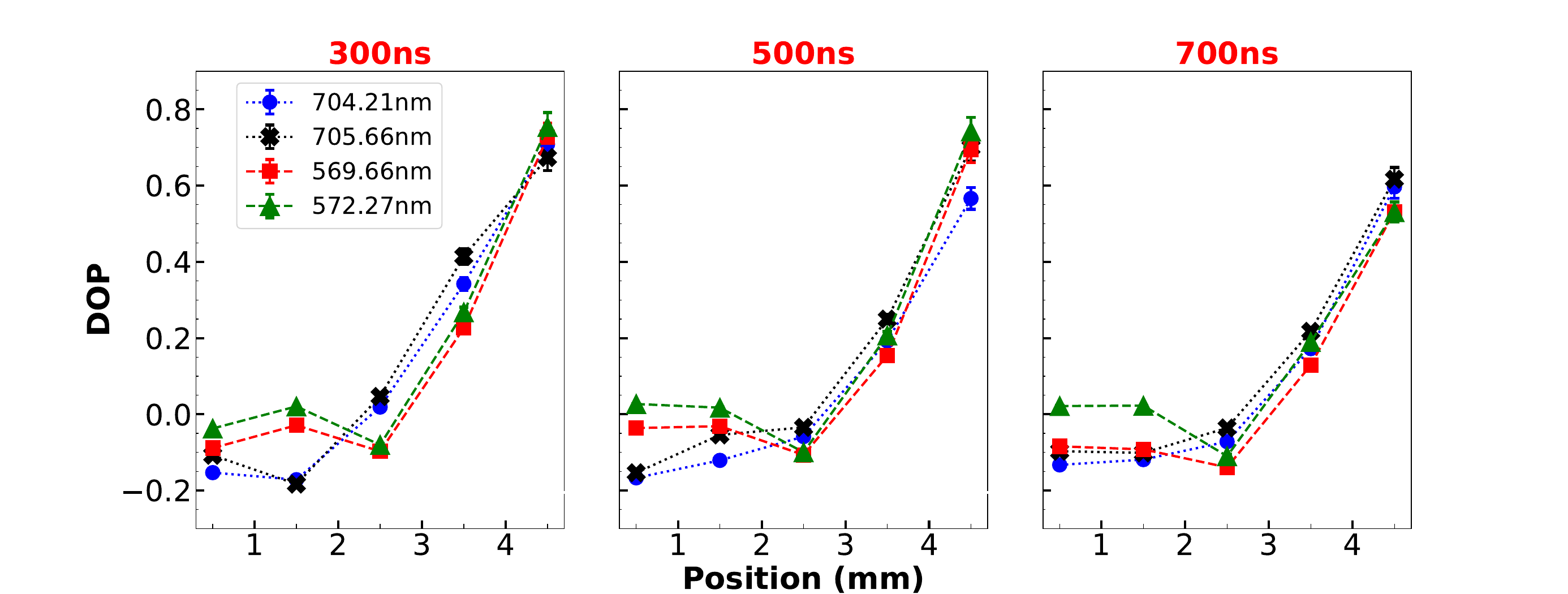}
	\caption{\label{fig:dop_704_569_572} DOP of  Al II- 704.21 nm and Al III- 569.66 nm and 572.27 nm) with distances from the sample at different times of plasma evolution.}
\end{figure}

\par

The behaviour of the DOP for the continuum and the line radiation appears interesting and requires further investigations to understand the underlying process behind this observed trend. As discussed earlier, the involvement of the self generated magnetic field\cite{SHARMA20073113,NAGLI} and its effect on the magnetic sublevels of a particular state can't be ruled out. Therefore, the DOPs of two Al III lines (572.27 nm ($^2$P$_{1/2}$ $\rightarrow$ $^2$S$_{1/2}$)and 569.66 nm ($^2$P$_{3/2}$ $\rightarrow$ $^2$S$_{1/2}$)) and two Al II lines (704.21 nm ($^3$P$_{2}$ $\rightarrow$ $^3$S$_{1}$) and 705.66 nm ($^3$P$_{1}$ $\rightarrow$ $^3$S$_{1}$)) under the same experimental conditions are compared. The ratio of $\pi$ and $\sigma$ transitions in the emission intensity varies for these lines. As can be seen from figure~\ref{fig:Jvalues}, the number of $\pi$ and $\sigma$ transition for Al III 572.27 nm and 569.66 nm lines are 1:1 and 1:2 respectively and for Al II 704.21 nm and 705.66 nm lines are 1:2 and 3:4 respectively.
Figure \ref{fig:dop_704_569_572} shows the variation of the DOP of these four lines with distance at three different delay times. At shorter distances from the sample, DOP is insignificant for the 572.27 nm emission line. However, DOP is higher for the other lines where the $\pi$ and $\sigma$ ratios are different from 1. This observation clearly gives the confidence about the proposed reason of self generated magnet filed \cite{SHARMA20073113} and its interaction with the magnetic sub levels\cite{Goto2021} as the reason for the observed anisotropy. However, observed flip and polarization at longer distances (2.5 mm and beyond) can't be related to the self generated magnetic field as it is expected to diminish as the plume propagates. We believe that the observed DOP for different charge states may be due to more than one process including role of the background pressure.

\par

 At a background pressure of 100 mbar, the plasma plume will face a substantial drag as it moves away\cite{harilal_press,Thomas_physD} and the plasma plume almost becomes stagnant. Also significant mixing between the background species and  plasma constituents are expected\cite{Garima_jaas} and the instabilities like Rayleigh-Taylor (RT) can be present \cite{SHARMA20073113,PRE_RT}.
Studies have shown that RT instability and pressure gradients at the plasma front can contribute to self generated magnetic field\cite{PRL_RT_Magnetic,PRA_Pr_Magnetic}. We believe that more experiments by varying the background gas pressure may shed more light into the underlying processes which will be attempted in future studies. 

\begin{figure*}[hbt]
	\includegraphics[scale=0.43,trim = {3.7cm 0 4.2cm 0},clip]{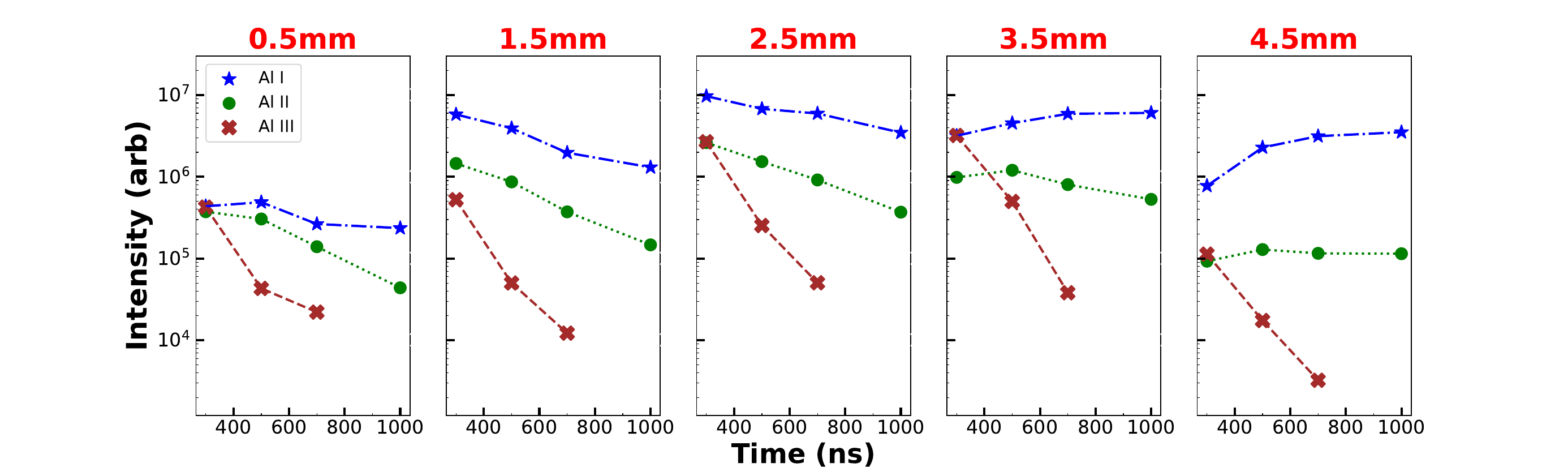}
	\caption{\label{fig:int_diffwave_corrected_log_mod_2} Time evolution of the total intensity (sum of both H and V) of the different ionic states of Al (Al I- 396.15, Al II- 704.21 and Al III- 569.66) at different distances from the sample.}
\end{figure*}

\par	
At this point, it would be interesting to look into the actual emission intensities of different charge states of Al  as it can provide some qualitative information about the recombination. For this, the intensities of H and V polarization were added together after subtracting the background count. Figure~\ref{fig:int_diffwave_corrected_log_mod_2} shows the temporal evolution of total line emission intensities of Al I (396.15 nm), Al II (704.21 nm), and Al III (569.66 nm) at different spatial points. At longer distances (3.5 mm and 4.5 mm), intensity of the neutral line increases with time, indicating that the recombination process possibly dominates for longer duration and distances. The decrease in emission intensity from higher charge states at longer distance also points towards the process of recombination. As DOP is primarily linked with different processes inside the plasma plume, it is obviously imperative to estimate plasma density and temperature at different distances and times.

\par

\begin{figure*}[hbt]
	\includegraphics[scale=0.43,trim = {4cm 0 4.2cm 0},clip]{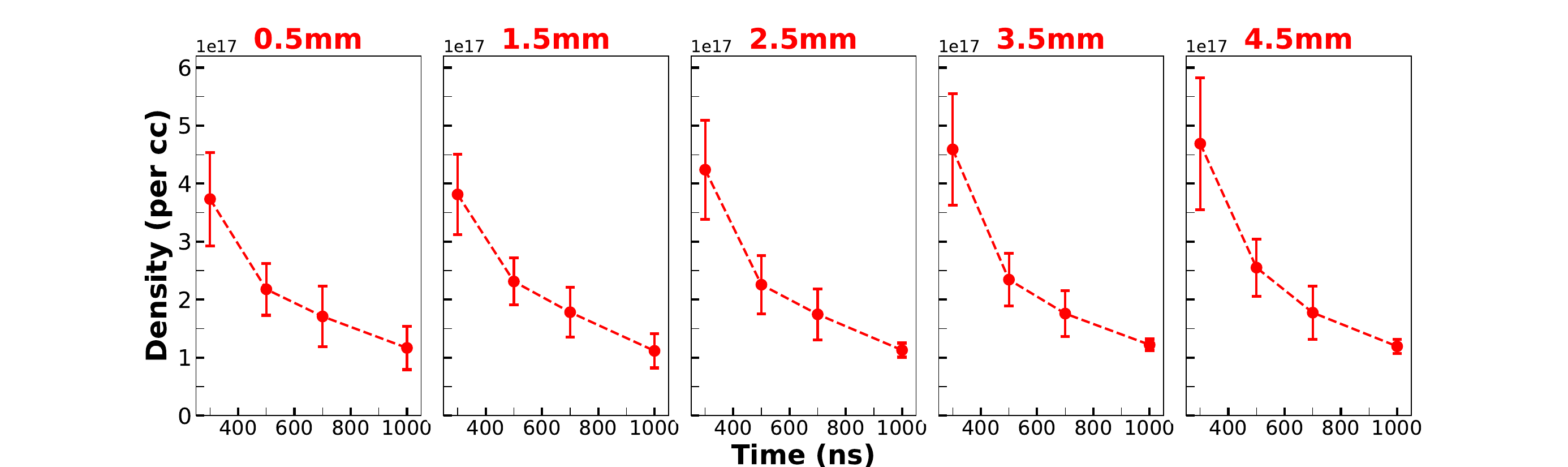}
	\caption{\label{fig:dens} Density variation of Al plasma with respect to time for different distances from the sample. }
\end{figure*}


At the early expansion stages of the laser produced plasma, Stark broadening is dominant among various broadening mechanisms\cite{densharilal,Mercadier2013}. So, the number density of electrons can be calculated from the Stark broadening parameter for a particular emission line by fitting the spectrum with Lorentzian function\cite{devia,JINTO_POP}. Contribution from instrumental broadening  (estimated using a low pressure calibration lamp) is subtracted from the line width as the instrumental broadening also has Lorentzian profile. The expression for the density in terms of the width of the spectral line for non-hydrogenic atoms is given by equation \ref{Stark_eq_small} \cite{densharilal}	
\begin{equation}
	\Delta \lambda_{1/2}=2\omega\left(\frac{N_e}{10^{16}}\right) \AA
	\label{Stark_eq_small}
\end{equation}
where $\lambda_{1/2}$ is the full width half maxima (FWHM) of the Lorentzian fit, $w$ is the electron impact parameter and $N_e$ is the electron number density ($cm^{-3}$). Various groups have studied the Stark width and electron impact parameters of emission from neutral and ionic lines of Al \cite{density559,density704,densal3} for a given temperature and density. We have calculated the electron density from Al II and Al III emission lines using the Stark width parameter as listed in table 2.

\begin{table}[h!]
	\caption{Spectral transitions and corresponding impact parameters used for estimating the electron density using Stark broadening}
	\centering
	\renewcommand{\arraystretch}{2}
	\begin{tabular}{ | p{1cm} | p{1.6cm} | p{2.5cm} | p{1.5cm} |}
		\hline
		Ionic state & Wavelength (nm)  & Impact parameter ($\AA $) , $N_e (cm^{-1})$ & Reference\\ \hline
		Al II & 559.33 & 0.38 , 0.1$\times 10^{17} $ & \cite{density559} \\ \hline
		Al II & 704.21 & 1.90, 1.00 $\times 10^{17}$ & \cite{density704} \\ \hline
		Al III & 452.92& 1.34, 1.00 $\times 10^{17} $ & \cite{densal3} \\ \hline
		Al III & 569.66 & 1.0, 1.00 $\times 10^{17} $ & \cite{densal3}\\ \hline
		
	\end{tabular}
	\label{table:dens}
\end{table}

Figure \ref{fig:dens} demonstrates the density variation with time at different locations in the plasma 
The values are averaged over the densities estimated using different lines of Al emission (452.92 nm, 569.66 nm, 559.33 nm and 704.21 nm). The statistical deviation, expressed as error bars in the figure, is rather small indicating good agreement in the estimate of densities even while using different emission lines. However, the density estimated from Al I is avoided since it yields a larger value indicating the possibility of self-absorption.
Further, the density is found to decrease with time as expected in the expansion of laser produced plasma. However, unlike freely expanding plasma (for high vacuum), the electron density does not decrease with distance probably due to the enhanced confinement of the plasma at high background pressure. In fact, the plasma density is increasing slightly along the propagation direction. 
Even at low background pressures, Coons et al\cite{Coons2011} have reported rather a constant spatial density profile. At higher background pressure, some of the earlier studies\cite{diwakar,MAL2021112839} have shown an increase in plasma density along the plume propagation direction as in this study.
\par
In the initial stages of expansion, laser-produced plasma is assumed to be in local thermodynamic equilibrium (LTE) due to its high density and moderately low temperature. Since the background pressure is 100 mbar in our case, the plasma is confined for longer duration, maintaining the LTE condition. The necessary criterion for plasma to be in LTE for an emission corresponding to energy $\Delta$E is given by the McWhirter criterion\cite{ARAGON2008893} (equation \ref{Mc_whiter_condition}), 
\begin{eqnarray}
	N_e (cm^{-3})  \geq 1.6\times 10^{12} T^{1/2}(\Delta E)^3
	\label{Mc_whiter_condition}
\end{eqnarray}
where $N_e$ is the number density of electrons, $T$ is the equilibrium temperature in Kelvin and $ \Delta E$ is the energy difference between levels involving the particular transition in eV. 

\par

Electron temperature can be estimated using several methods. Boltzmann plot method is one among them, which is widely used. However, accuracy in estimated temperature depends on the separation of upper state energies of the involved transitions. In our case, as can be seen in the table \ref{table:em_param},  the upper state energies of recorded spectral transitions of Al II do not differ much and hence, can have large errors in the estimated values. Thus, the temperature is estimated using the line intensities of successive ionic states, which is considered as more accurate than Boltzmann plot method\cite{grieim,densharilal}
The dependence of the ratio of intensities of successive ionized states is given by equation \ref{Temp_int_R}

\begin{align}
	\frac{I'}{I} &=\frac{f'g'\lambda^3}{fg\lambda'^3}(4\pi^{3/2}a_0^3N_e)^{-1}\bigg(\frac{k_BT_e}{E_H}\bigg)^{3/2} \label{Temp_int_R} \\
	&\times exp\left(\frac{-(E'+E_{\infty}-E-\Delta E_{\infty})}{k_BT_e}\right) \notag		
\end{align}

where $\lambda, f, g, I $ and $ E $ are the wavelength, oscillator strength, statistical weight, line intensity, and upper level energy respectively for the lower ionic state. Similarly, $ \lambda', f', g', I' $ and $ E' $ are for higher ionic state. $E_{\infty}$ is the ionization energy of the lower ionic state and $\Delta E_{\infty}$ is the correction to the ionization energy. $E_H$ is the ionization energy of hydrogen atom, $a_0$ is the Bohr radius and $k_B$ is the Boltzmann constant. $N_e (m^{-3})$ and $T_e$ (Kelvin) are the density and temperature of electrons in the plasma. Equation \ref{Temp_int_R} can be used to estimate the electron temperature from the line intensity ratio between Al II and Al III lines if an accurate estimate of plasma density is available.  As the statistical variation in calculated densities from  different spectral transitions is reasonably small (figure \ref{fig:dens}), it is safe to use for temperature estimation. The temperature estimated using this method shows that a maximum value around 2.1 $\pm$ 0.3 eV at 300 ns which eventually falls to 1.6 $\pm$ 0.2 eV at 700 ns at a given position. However, we observe that the temperature does not vary significantly along the propagation direction, similar to the observation of Coons et al \cite{Coons2011}. The consistency in temperature with distance can be due to the confinement of plasma at high background pressure and plasma thermalization\cite{diwakar}.

\par
\begin{figure}[hbt]
	\includegraphics[scale=0.21,trim = {0.55cm 0 4.5cm 1cm},clip]{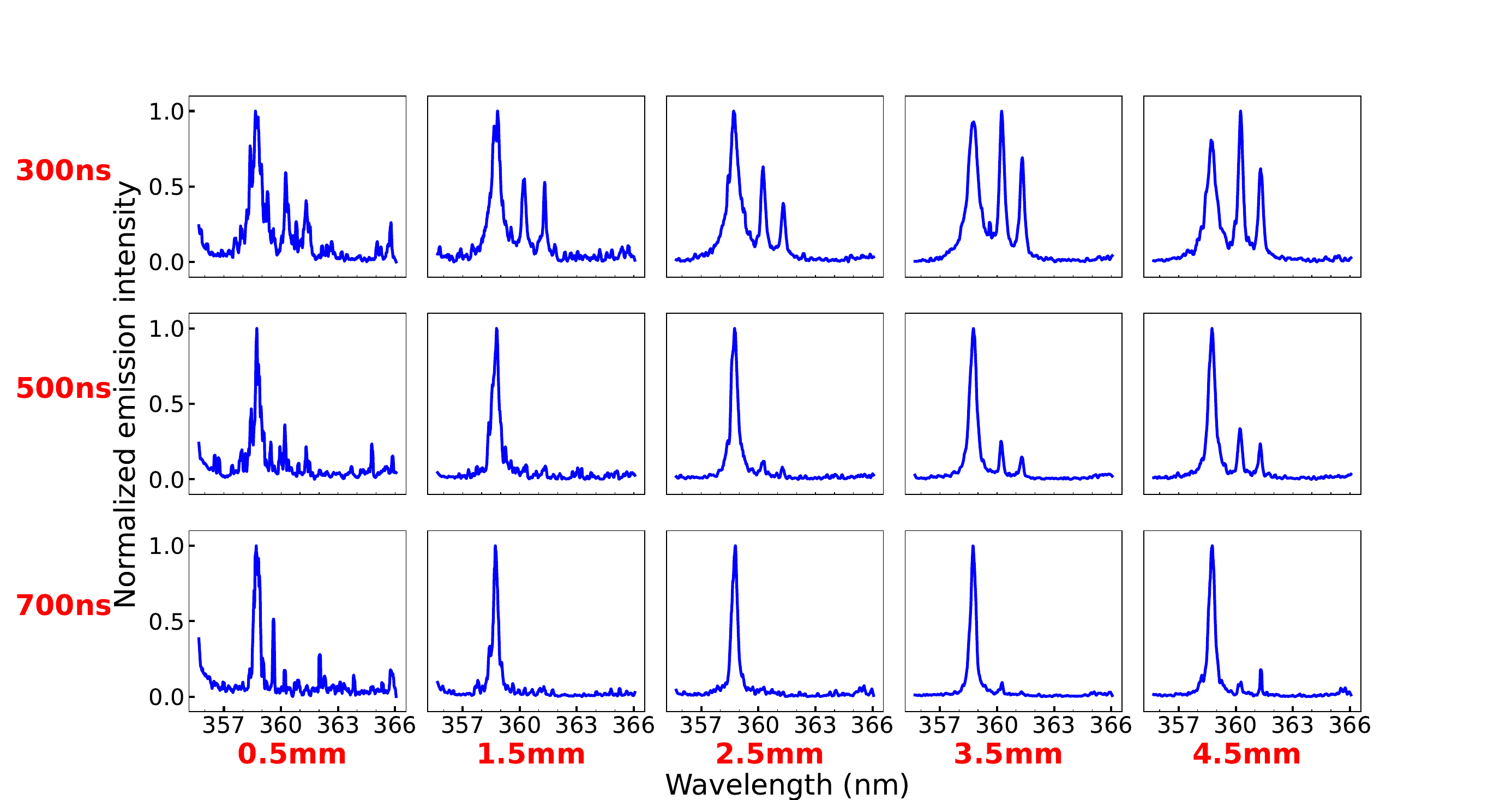}
	\caption{\label{fig:360_ch1} Emission spectra of Al II line at 358.69 nm and Al III lines at 360.16 and 361.23 nm indicating rather constant temperature and density profile}
\end{figure}

To further validate the observed flat spatial profile for temperature and density the spectra of Al II and Al III emissions given in figure~\ref{fig:360_ch1} can be checked. From the figure the spectral behaviour for Al II and Al III at a given time appears rather same along the propagation direction, indicating the estimated consistent temperature and density profiles are indeed correct. Additionally, falling of density with time at a particular position also can be seen from the figure.

\par

In addition to the self generated magnetic fields, the causes of anisotropic emission has been attributed to spatial anisotropy of electron distribution\cite{Hammond1989,Wolcke_1983} , properties of states pumped\cite{NAGLI}, gain cross section on the reabsorption \cite{PRL_Re_Abs}, self generated electric field due to double layer\cite{bulgakova} and recombination nature of the plasma\cite{PRE_DOP_RC}.  Anisotropic ion emission during ablation\cite{lucas_euv} can also be the cause of considerably higher DOP in case of ions as compared to neutrals. Any of these factors can act along with the self generated magnetic field resulting the observed flip in polarization. However, it is difficult to conclude which particular process is responsible for the observed behaviour.

 Hammond et al\cite{Hammond1989} and Wolcke et al\cite{Wolcke_1983} have shown that the DOP due to resonant impact excitation by electron causes polarization and its flip based on the energy of the electron. 
However, In a thermalized plasma, as seen from the estimated temperate and density, the possibility of anisotropic electron distribution is rather unlikely. Recombination can expected to be there but we observe the density is not changing significantly from 3.5 mm to 4.5 mm, where large variation in DOP is observed.  This points to the fact that recombination may not be responsible for it.
We have also investigated the effect of laser polarization on DOP by changing the polarization of the laser pulse used for ablation and observed that the DOP has not been affected by the laser polarization, which is consistent with the findings of earlier investigations\cite{kieffer1992,spatial_pps}.
\par

Fine understanding towards the cause of observed anisotropy can be quite challenging and may need further experimental and theoretical studies. 
Nonetheless, present study clearly demonstrates that a complete spatio-temporal study involving various transitions is an important aspect in enhancing the understanding towards the mechanistic aspect. 	

\section{conclusion}
\label{sec:conclusion}

In the present study, a precise imaging system has been used to study the spatial and temporal variations of anisotropy in laser-produced Al plasma with a Wollaston prism for separating the polarizations. The results show an interesting behaviour in DOP of the emissions depending on the spatial location. The DOP close to the sample shows a dominant V polarization which eventually changes to H as the plume propagates, an observation reported for the first time in laser-produced plasma. Moreover, the observed anisotropy depends on the charge state of the emitting species. In the early stages and close to the sample, the continuum emission predominantly shows a V polarization, indicating the role of non-thermal electrons due to inverse-Bremsstrahlung absorption. Plasma temperature and density appear to have a flat spatial profile indicating rather thermalized and confined plasma. The observed polarization of line emission close to the sample can be due to the self-generated magnetic field. In addition to the self-generated magnetic field, several competing processes may be taking place which contributes to the observed behaviour.  The reason for the flip in the DOP around 2.5 mm from the sample and a dominant H polarization away from the sample appears to be due to high background pressure. The present study is able to rule out a few possibilities widely believed as the cause of polarized emission, but a more in-depth experimental and theoretical studies need to be carried out to have further insight into the observed anisotropic nature of emission from laser produced plasma.

\section{References}


%


\end{document}